# Eco-Mobility-on-Demand Fleet Control with Ride-Sharing

Xianan Huang, Boqi Li, Huei Peng, Joshua A. Auld, Vadim O. Sokolov

*Abstract*—Shared Mobility-on-Demand using automated vehicles can reduce energy consumption and cost for future mobility. However, its full potential in energy saving has not been fully explored. An algorithm to minimize fleet fuel consumption while satisfying customers' travel time constraints is developed in this paper. Numerical simulations with realistic travel demand and route choice are performed, showing that if fuel consumption is not considered, the Mobility-on-demand (MOD) service can increase fleet fuel consumption due to increased empty vehicle mileage. With fuel consumption as part of the cost function, we can reduce total fuel consumption by 7% while maintaining a high level of mobility service.

## I. INTRODUCTION

Ground transportation consumes 26.5% of the world energy in 2016 [1]. In 2014, 3.1 billion gallons of fuel and 6.9 billion hours of time are wasted due to congestion [2]. Mobility-on-demand (MOD) services such as Uber and Lyft have brought significant changes, especially in urban areas with dense population. When multiple passengers share the same vehicle (e.g., Lyft Line and UberPOOL), the number of vehicles parked and on the road will reduce, which in turn can reduce congestion and energy consumption. Intelligent transportation techniques enable smarter planners to reduce travel time and fuel [3], and the potential has not been fully explored.

Although eco-driving and eco-routing concepts have been proposed to reduce fuel consumption and emission at the operation level, the major cause for fuel consumption increase is the additional travel demand such as currently underserved population (2% ~ 40%), travel mode shift (~3.7%), and empty vehicle mileage (0%~11%) as pointed out by a recent study on potential impact on fuel consumption of connected automated vehicle (CAV) technologies [4]. Thus, ride-sharing is proposed to reduce fuel consumption directly at the travel demand level [5] and has the potential to reduce vehicle mileage traveled by 12% [6]. However, the current fleet assignment of MOD from literature are either travel time oriented [7–12] or fleet sizing oriented [13–16], and the benefits of fuel-saving are mainly due to reduced trips [17]. The full potential in fuel-saving by including trip-level techniques such as eco-routing or minimizing total fleet fuel consumption in assignment optimization was not addressed in the literature.

Control of MOD fleet has been studied extensively to minimize customers' travel time. The fleet assignment problem falls in the category of dynamic Vehicle Routing Problem (VRP) [18] in the demand-vehicle network, which is a generalization of Traveling Salesman Problem (TSP). The problem is typically formulated as an integer programming problem. Several studies developed algorithms to find the exact solution [19–21]. However, due to the NP-hardness of VRP [22] and the large problem size, the centralized matching problem is hard to solve directly [23]. Thus, heuristic algorithms such as Genetic/Evolutionary algorithms combining insertion algorithm [24, 25] and bee colony optimization [26] are applied to find a suboptimal solution. Decomposition-based algorithms focus on reducing the problem size either spatially [27] or use Lagrange relaxation [28] to combine multiple smaller TSP into the master VRP, thus the solving process is accelerated due to the reduction of problem size and parallelization. Recently, [8] demonstrated that the current travel demands for taxis in New York City could be fulfilled by a MOD fleet 15% the size of the existing fleet. A data-driven approach is used to further improve the quality of the solution by considering future demands [7]. [29] developed a simulation optimization (SO) framework using continuous approximation as a metamodel to improve the computational efficiency. Other aspects of MOD systems were also explored. A privacy-preserving algorithm was developed [30] to protect the location information of passengers without a significant performance hit. Continuous approximation [31] is used to study the dynamics of the fleet and the influence of large fleet to congestion as well as the fleet routing problem in a congested network [10, 32]. The trade-off between the customers' travel time requirements and the system operator's cost is studied [33], where the system operator's cost is modeled as the time each vehicle spends in operation. Using fuel consumption as the cost, [34] solved green VRP without considering travel time constraints, thus cannot be applied to MOD system directly. As far as we know, none of the existing work considers fuel consumption when designing the controller, which is a core element in reducing the operation cost of the MOD service provider.

To include fuel consumption in the objective function and integrate MOD fleet control with the recent eco-routing [3] concept, we developed a fleet control algorithm based on the work in [8] where the customers' wait time and travel delay time are modeled as constraints. We propose a MOD fleet control algorithm, Eco-MOD, to minimize the fleet operation cost (fuel consumption) while satisfying the customers' travel time constraints. In our study, travel demands generated by

Manuscript received XXX,XX, 2019. The research is supported by U.S. Department of Energy under the award DE-EE0007212.

X. Huang, B. Li, H. Peng are with the Department of Mechanical Engineering, University of Michigan, Ann Arbor, MI 48109, USA (e-mail: xnhuang@umich.edu, boqili@umich.edu, hpeng@umich.edu).

J. Auld is with the Energy Systems Division, Argonne National Laboratory, Lemont, IL. 60439, USA (e-mail: jauld@anl.gov)

V. Sokolov is with the Systems Engineering and Operations Research Department, George Mason University, Fairfax, VA, 22302, USA (e-mail: vsokolov@gmu.edu).

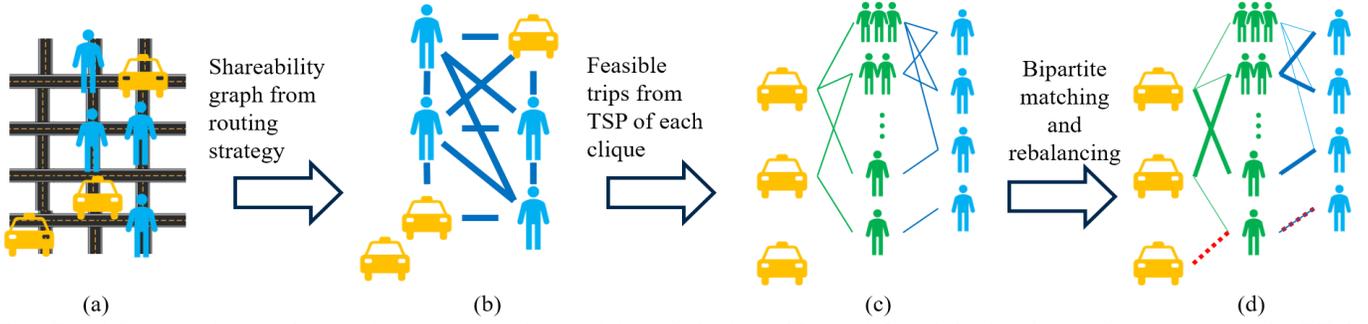

Fig. 1 Travel demand assignment framework: (a) system receive travel demand; (b) shareability graph formulation based on routing strategy; (c) solve TSP for each clique in the shareability graph to get all feasible trips; (d) assign trips to vehicles and assign ignored customers to idling vehicles for rebalancing, with thick solid line indicating feasible trip assignment and dashed line showing passive rebalancing assignment

POLARIS [35], a mesoscopic agent-based transportation model, are calibrated with data from the Safety Pilot Model Deployment (SPMD) project [36] to generate the origins and the destinations of the customers. To evaluate the performance of Eco-MOD under realistic transportation environment, we developed a microscopic traffic simulator using Simulation of Urban Mobility (SUMO) [37] and performed a case study using the integrated model. In this study, the data-driven fuel consumption model from [3] is used to estimate fuel consumption cost for fleet assignment and eco-routing. The fleet is assumed to be homogeneous, and consists only internal combustion engine (ICE) vehicles with powertrain parameters from [3].

The main contributions of this work are: 1) a MOD fleet control algorithm which minimizing fleet fuel consumption directly while satisfying customer travel time constraints; 2) a simulation framework for MOD system with microscopic simulation from SUMO and demand generation from POLARIS; 3) demonstrating the importance of including fuel consumption in the optimization cost function to reduce fleet operating cost.

The rest of this paper is organized as follows: Section 2 presents the formulation of fuel-efficient ride-sharing fleet optimization. Section 3 presents the simulation framework to evaluate the performance of the MOD fleet. Section 4 presents the cost configurations for numerical simulation. Section 5 presents the simulation results. Conclusions and future work are given in Section 6.

## II. Fleet Assignment

### A. Travel Demand Assignment

Our fleet control algorithm is based on the graph decomposition method proposed in [8]. The algorithm can solve the trip matching and routing problem for ride-sharing for thousands of vehicles and customers fast enough for real-world implementation. We further improve the algorithm to take knowledge of fuel consumption as the fleet operation cost. The framework to solve the fleet control problem is summarized in Fig. 1.

We reproduce the work in [8] by assuming the road network is static and solving all optimal routes considering travel time and fuel consumption offline. Including dynamic road network information is considered as part of our future work. The trip assignment algorithm is based on a shareability graph (Fig. 1.b). The graph is defined as an undirected graph with nodes defined as customers and vehicles. The constraints for each customer consist of wait time and delay time. Wait time is defined as the time between the customer travel request and time of pickup. Delay time is defined as the difference between planned travel time and the shortest travel time after pickup, which is from the minimum cost routing solution from origin to destination. In [8], travel time is used as the cost, and we expand the routing cost to include fuel consumption. An edge exists between two customers if a virtual vehicle can depart from the origin of one of the customers and fulfill the travel demands of both customers without violating travel time constraints. An edge exists between a vehicle and a customer if the demand can be served by the vehicle without violating travel time constraints. Then a necessary condition for a trip to be feasible is that the customers of the trip can form a clique with one vehicle present in the shareability network. A clique is a subgraph such that every node is connected to every other node within the same clique. It is noted that the cliques do not need to be maximum in the shareability graph. The cliques in a graph can be found with the Bron-Kerbosch algorithm [38] with worst-case time complexity $O(dn3^{d/3})$ where $n$ is the number of nodes and $d$ is degeneracy of the graph, which is a measure of sparseness. In this way, instead of evaluating the cost for every possible combination of customers and vehicles, one can solve single-vehicle-multiple-customer problems modeled as TSP for every clique, a necessary condition for a trip to be feasible.

Trip scheduling for each clique is a traveling salesman problem with pickup and delivery. The problem can be solved with multiple algorithms. If the number of customers is small, (e.g., less than 5), the exact solution can be found by Dynamic Programming in less than 1 sec on a standard desktop computer. Heuristic-based algorithms such as T-share [39] can be used to find the solution if the problem size is large. In our numerical study, the vehicle capacity is set at 4, and Dynamic Programming is used to find the exact solution. The states are defined as

$$\boldsymbol{\delta}_t = [\delta_{1,t}^P, \cdots, \delta_{i,t}^P, \cdots, \delta_{N,t}^P, \delta_{1,t}^D, \cdots, \delta_{i,t}^D, \cdots, \delta_{N,t}^D] = [\boldsymbol{\delta}_t^P, \boldsymbol{\delta}_t^D]$$

where $\delta_{i,t}^P$ and $\delta_{i,t}^D$ are indicator variable for pickup location and drop-off location of customer i at step t respectively, the value is 1 if the location has been visited and 0 otherwise. If two customers have the same pickup or drop-off locations, we assign individual variables for them, but define the transitional cost as 0. $N$ is the total number of customers in the clique. The problem is to find the optimal trajectory to travel from the initial state, which is $\boldsymbol{\delta}_0 = \{0\}_1^{2N}$, to the terminal state, which is $\boldsymbol{\delta}_T = \{1\}_1^{2N}$. The constraints are defined as follows:

$$\delta_t^D - \delta_t^P \geq 0, \forall t \quad (1)$$

indicating that the drop-off locations are visited after the pickup locations of each customer.

$$\sum_i \delta_{i,t}^P - \delta_{i,t}^D \leq V_c, \forall t \quad (2)$$

where $V_c$ is the capacity of the vehicle, indicating the number of customers onboard should not exceed the capacity of the vehicle. The continuity constraint is defined as

$$\|\boldsymbol{\delta}_{t+1} - \boldsymbol{\delta}_t\| = 1, \forall t \quad (3)$$

indicating that only 1 pickup/drop-off happens for each state. If the objective for fleet assignment is minimizing wait time and delay time of customers, the transitional cost is defined as

$$g(t, t+1) = \sum_i T_{t,t+1}\left((1 - \delta_{i,t}^P) + w_D(\delta_{i,t}^P - \delta_{i,t}^D)\right) \quad (4)$$

where $T_{t,t+1}$ is the travel time from location at $t$ to $t+1$, $w_D$ is the weighting parameter between wait time and on-vehicle travel time of the customers as in [8]. In our eco-MOD framework, the fuel consumption of traveling between locations associated with the states is used as the transitional cost. The objective of the TSP is minimizing the sum of the transitional costs from the initial state to the terminal state

$$J_{TSP} = \sum_{t=0}^{T-1} g(t, t+1) \quad (5)$$

where $J_{TSP}$ is objective of the TSP step. A trip is feasible if the wait time and delay time constraints are satisfied for all customers in the clique. After all feasible trips were found through solving the TSP for all cliques (Fig. 1.c middle column), the optimal trip assignment problem can be formulated as a bipartite matching problem and solved through Integer Linear Programming (ILP).

The cost of a trip is denoted as $c_t^i$ for trip $i$. The states of the system are $\delta_t$ which is the indicator variable for the trip/clique and $\delta_c$ which is the indicator variable for the customer. If at an assignment instant, there are $m$ feasible trips from the TSP step and $n$ customers, then $\delta_t = \{\delta_t^i \in \{0,1\}, i \in \mathbb{N}, 1 \leq i \leq m\}$ and $\delta_c = \{\delta_c^i \in \{0,1\}, i \in \mathbb{N}, 1 \leq i \leq n\}$. $\delta_t^i$ is 1 if trip $i$ is selected and $\delta_c^i$ is 1 if customer $i$ is assigned. The objective function is

$$\sum_{i=1}^m c_t^i \delta_t^i + \sum_{i=1}^n D(1 - \delta_c^i), \quad (6)$$

where $D$ is the penalty for unserved customers. In the original fleet control problem[8], a weighted sum of total wait time and delay time of each trip is used as cost, and in eco-MOD framework, the total fuel consumption is used as the cost. The constraint for the vehicle is that each vehicle can only serve one trip

$$\sum_{i=1}^m a_j^i \delta_t^i \leq 1, \forall j, \quad (7)$$

where $a_j^i$ is the indicator variable for vehicle $j$ and trip $i$, $a_j^i = 1$ if vehicle $j$ can serve trip $i$. The constraint for the customer is that a customer is either assigned or ignored

$$\sum_{i=1}^m b_j^i \delta_t^i + (1 - \delta_c^j) = 1, \forall j, \quad (8)$$

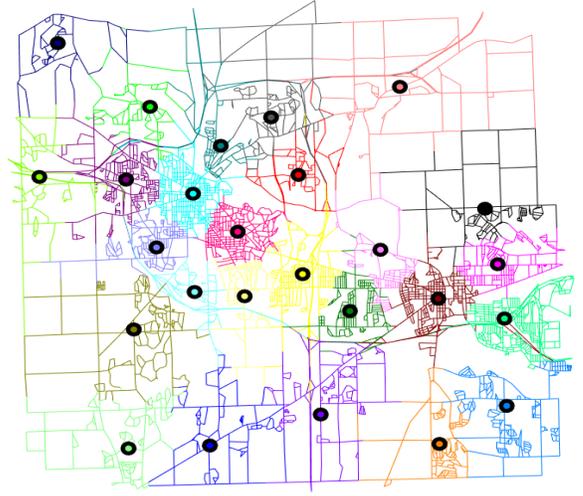

Fig. 2 Partitioned road network

where $b_j^i$ is the indicator variable for customer $j$ and trip $i$, $b_j^i = 1$ if customer $j$ can be served by trip $i$. With linear constraints and the objective function, the trip assignment problem is an integer linear programming (thick lines in Fig. 1.d). Since all candidate trips are feasible from construction, the travel time constraints are satisfied. For online optimization, we follow [8] to keep a pool of customers, and a customer is removed from the pool if it's picked up by vehicle or cannot be served within the time constraint. If a customer is not able to be served within the travel time constraint, a vehicle from the idling fleet is assigned to serve the customer with the minimum wait time as the objective (dashed lines in Fig. 1.d), which is referred to as passive rebalancing and different from the active idling fleet rebalancing trips introduced in the next section. Gurobi [40] is used to find the solution of ILP. The optimization problem is solved every assignment interval reacting to new travel requests.

*B. Active Idle Fleet Rebalancing*

Since there is a mismatch between the trip origin distribution and the trip destination distribution, the idling vehicles tend to build up with the trip destination distribution, which would increase the expected wait time of new customers. To mitigate this effect, the idling vehicles should be relocated according to the trip origin distribution to reduce the expected wait time for future travel demands. The road network is partitioned and trip origin distribution is modeled using the approach proposed in [41], the partitioned road network is shown in Fig. 2. After the road network is partitioned, the trip origin distribution and idling fleet distribution can be described using random variables following a categorical distribution. The objective of fleet rebalancing is to minimize the difference between these two distributions. Assuming known trip origin distribution and expected customer departure rate, the problem is formulated as an integer programming with quadratic objective function and linear constraints. To reduce the size of integer programming, we only consider the case that idling vehicles being relocated to immediate adjacent partitions.

The decision variables are defined as indicator variables of relocating trips associated with each idling vehicle, $\mathcal{T} = \{t_{ij}, \forall i \in \mathcal{V}_I, \forall j \in adj(v_i)\}$, where $\mathcal{V}_I$ is the set of idling vehicles, $adj(v_i)$ is the set of adjacent partitions of the idling vehicle $v_i$. $t_{ij} = 1$ if idling vehicle $i$ is assigned to be relocated to adjacent partition $j$, and otherwise is 0. For simplicity, we assign destinations of relocating trips at the corresponding partition center. The objective function is estimated using a planning horizon $T$, and we only consider assignment at the current assignment step. The objective function is defined as

$$\min_{t_{ij} \in \mathcal{T}} (1 - w_c) \sum_k \sum_{\tau=1}^{T} \left\| \frac{n_k^\tau}{N_\tau} - o_k \right\|^2 + w_c \sum_{t_{ij} \in \mathcal{T}} C_{ij} t_{ij}, \quad (9)$$

where $N_\tau$ is the normalization constant, $o_k$ is the density function of trip origins associated with partition $k$, $C_{ij}$ is the traveling cost associated with relocating trip $t_{ij}$, and fuel consumption is used here, $w_c$ is weighting parameter between relocating cost and balancing objective, $n_k^\tau$ is the estimated number of idling vehicles within partition $k$ at time $\tau$. Assuming that customers' departure process in each partition follows a Poisson process, the expected number of idle vehicles in each partition is given by

$$n_k^\tau = \sum_{t_{ij} \in \mathcal{T}} t_{ij}^{k,\tau} + d_k^\tau - \tau \gamma \lambda_k, \quad (10)$$

where $t_{ij}^{k,\tau}$ is trajectory indicator of rebalancing trip $t_{ij}$, $t_{ij}^{k,\tau} = 1$ if the vehicle is within partition $k$ at time $\tau$, which can be estimated as a function of $t_{ij}$ using estimated travel times, $d_k^\tau$ is the amount of arrival vehicles within partition $k$ up to time $\tau$, $\gamma$ is a discount factor to account for the ratio of trips being shared, $\gamma \leq 1$, $\lambda_k$ is the expected customer departure rate of partition $k$. The normalization constant is given by

$$N_\tau = \sum_k n_k^\tau = N_{idle} + \sum_k d_k^\tau - \tau \gamma \sum_k \lambda_k, \quad (11)$$

where $N_{idle}$ is the amount of idling vehicles. In addition to the trips to other partitions, virtual trips that the vehicle stays at the same location are also generated with the destination assigned to be the vehicle's current location. Thus the constraint is that each vehicle is assigned to one relocating trip.

$$\sum_j t_{ij} = 1, \forall i, \quad (12)$$

Due to the large size of the problem, instead of solving the integer programming exactly using the branch and bound algorithm, we solve the continuous relaxation of the original problem by relaxing $t_{ij}$ as a real number between 0 and 1. The integer solution is then obtained by randomized rounding [42]. After solving the relaxed problem, the assigned trips are selected by random sampling using the optimum $t_{ij}$ as the density function. The idling fleet rebalancing step is integrated with the MOD assignment framework by assigning idling fleet relocating after the reactive rebalance step to serve customers whose travel time constraints cannot be satisfied by the regular assignment. The optimization is solved with Gurobi [40]. The optimization is solved repeatedly every assignment interval based on the current status of the fleet.

## III. TRAFFIC SIMULATOR

POLARIS is an agent-based traffic simulation software developed by the Argonne National Lab [35] focusing on travel demand and mesoscopic traffic simulations. Travel demand is generated using the ADAPTS (Agent-based Dynamic Activity

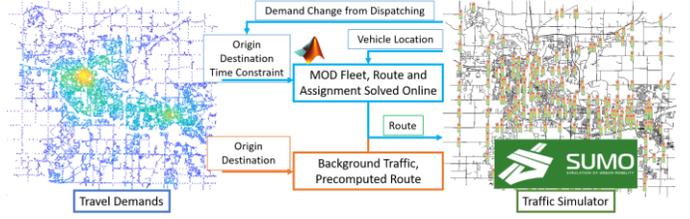

Fig. 3 Transportation Simulation Framework

Planning and Travel Scheduling) model in POLARIS, which formulates the activity planning of individuals as a dynamic model [43]. The demand model is calibrated by the Argonne National Lab using the dataset from the Safety Pilot Model Deployment project, which is real data from the city of Ann Arbor. The data is aggregated over 5 months, from May 2013 to October 2013. However, as a mesoscopic simulator, POLARIS's ability to simulate individual vehicle's dynamics is limited. Thus, POLARIS is used as travel demand generator, with which 110,000 trips are generated from 17:00 to 19:00 and a microscopic transportation simulator, Simulation of Urban Mobility (SUMO) [37] is used for verification and to generate individual vehicle trajectories with speed changes. At each simulation step, the locations of MOD fleet vehicles are updated using SUMO.

SUMO is an open-source microscopic traffic simulator with the ability to generate realistic speed profile. In the simulations, the traffic demand and route choice are calibrated using data from SPMD and the demand generated by POLARIS is treated as a prior. A random subset of demands are assumed to be served by the MOD fleet. We assume the ratio of the MOD customers to the total demand is fixed. The fleet size is assumed to be fixed and ride-sharing is allowed. A fleet controller implemented in Matlab is used to control the route choice of the MOD vehicles using the TraCI4Matlab package [44], while the route of background traffic is precomputed and the route choice is from the calibration process introduced in this section. The simulation framework is summarized in Fig. 3.

The model is calibrated using the measured average speed from SPMD. In the calibration process, we solve for the route choice and travel demand distribution to minimize the difference between simulated and measured average speed. Demand generated by POLARIS is used as a prior for demand distribution estimation. The microscopic model parameters including the car-following model and the lane-change model are obtained from [45]. In the simulation framework, we only consider passenger cars. To estimate demand distribution given average speed measurement, we use a data-driven approach to model the relationship between the vehicle density and the average travel speed for links in SUMO, which is used to estimate the expected flow rate at each link given the measured average speed. A second-order polynomial is used when the density is below critical density for simplicity. When the vehicle density is higher than the critical density $\rho_{critical}$, we assume the average speed is a constant.

$$\overline{v_n} = \begin{cases} \epsilon & \rho \geq \rho_{critical} \\ \alpha_2 \rho^2 + \alpha_1 \rho + \alpha_0 & otherwise \end{cases}, \quad (13)$$

where $\overline{v_n}$ is normalized average speed, defined as average speed normalized by the free-flow speed. $\rho$ is the vehicle density at each link, $\epsilon$ is the normalized average speed when the vehicle density is greater than the critical density. Flow rate, vehicle density and average follow are related by

$$q = N\rho\bar{v}, \quad (14)$$

where $\bar{v}$ is the average speed, $q$ is flow rate, and $N$ is the number of lanes. Given the measured average speed from SPMD, the flow rate $\widehat{q_{SPMD}}$ is estimated. To estimate the travel demand and route choice, we assume the drivers follow the shortest distance or empirical shortest time route. Under the assumption that the system has reached steady state, given the flow rate between origin-destination pair $q_{od}$, the flow rate for each link is given by

$$q_l = \sum_k \left( q_{od}^{k,d} i_{od,d}^{k,l} + q_{od}^{k,t} i_{od,t}^{k,l} \right), \quad (15)$$

where $i_{od,d}^{k,l}$ and $i_{od,t}^{k,l}$ are indicator variables representing that link $l$ is used by OD pair $k$ following shortest distance route and empirical shortest time route respectively, $q_{od}^{k,d}$ and $q_{od}^{k,t}$ are the flow rate for OD pair $k$ following the shortest distance route and empirical shortest time route respectively. The OD flow is modeled using the partitioned road network from [41]. The total flow rate for OD pair $k$ is given by

$$q_{od}^k = q_{od}^{k,d} + q_{od}^{k,t}, \quad (16)$$

The objective of the calibration is to minimize the difference between the simulated flow rate and the estimated flow rate using the data-driven model from SPMD.

$$\min_{q_{od}^{k,t}, q_{od}^{k,d}} \sum_l \left\| q_l - \widehat{q_{l,SPMD}} \right\|^2 + \sum_k \psi \left\| q_{od}^k - q_{od,POLARIS}^k \right\|^2, \quad (17)$$

where $\widehat{q_{l,SPMD}}$ is estimated link flow rate from SPMD using the data driven model, $q_{od,POLARIS}^k$ is OD flow rate from POLARIS. $\psi$ is the weighting parameter between flow rate approximation and the regularization term. Assuming that the OD flow rate follows Gaussian distribution, the objective function is equivalent to the maximum-a-posterior estimation of the expected OD flow rate using POLARIS OD flow rate as prior. Assuming the total flow rate follows the total flow rate generated by POLARIS, we have the constraint

$$\sum_k q_{od}^k = \sum_k q_{od,POLARIS}^k, \quad (18)$$

The objective function is quadratic in OD flow rate and the constraints are linear, thus the optimization problem is convex. The quadratic program is solved using Gurobi. Shortest distance route and shortest time route are generated offline, and the percentage of drivers following the shortest distance in each OD flow is obtained by solving (17). To generate the empirical shortest time route, we use measured average speed, and for links with inadequate data, we assume that the average speed equals to the posted speed limit. We assume that the drivers follow real-time shortest time routes are uniformly distributed in the road network and the ratio is estimated by simulation. Also, we assume that the average speed on each link is normally distributed. The real-time routing ratio with the maximum likelihood of the average speed is selected as the optimum value. If the variances of average speed distribution are equal for all links in the network, this is equivalent to minimize the squared error between simulated and measured mean value of average speed.

TABLE I MOD FLEET ASSIGNMENT STRATEGY CONFIGURATION SUMMARY

|   | Assignment Cost | Assignment Routing Strategy | Passive Rebalance Routing Strategy |
|---|---|---|---|
| 1 | Trip Time | Fastest Routing | Fastest Routing |
| 2 | Trip Time | Fastest / Eco Routing | Fastest / Eco Routing |
| 3 | Trip Time | Eco Routing | Fastest Routing |
| 4 | Trip Time | Eco Routing | Eco Routing |
| 5 | Fleet Fuel | Fastest Routing | Fastest Routing |
| 6 | Fleet Fuel | Fastest / Eco Routing | Fastest / Eco Routing |
| 7 | Fleet Fuel | Eco Routing | Fastest Routing |
| 8 | Fleet Fuel | Eco Routing | Eco Routing |
| 9 | - | Shorest Distance/ Fastest | - |

## IV. ECO-MOD COST CONFIGURATIONS

Two levels of strategies can be used by the MOD fleet to reduce fuel consumption. At the trip assignment level, the objective function for the fleet assignment of the feasible trips can be the total fleet fuel consumption instead of the sum of individual's wait time and delay time as defined in the original fleet assignment problem [8]. However, for the assignment of the passive rebalance fleet, where the main objective is to serve the customers whose travel demand cannot be satisfied within the travel time constraints, we minimize their wait time when assigning the idling vehicles to the passive rebalancing trips. At the trip execution level, the routing strategy can be either shortest-time routing or eco-routing, and the corresponding routing cost is applied for the trip assignment. To assess the fuel-saving benefit of the two levels, eight test configurations are defined based on combinations of the cost function and the routing strategy. For all configurations, the passive rebalancing trips are assigned to minimize the travel time under the corresponding routing policy, while eco-routing is used for active rebalancing trips. The configurations are summarized in TABLE I, where the assignment of the feasible trips is denoted as assignment, and the assignment of the passive rebalance trips is denoted as passive rebalance. Configuration 9 is the baseline where personal vehicles are used, and the routing strategy is from the calibration results of the traffic simulator discussed in Section III.

As shown in TABLE I, configurations 1-4 weigh more on travel time, while configurations 5-8 weigh more on fuel consumption. The travel time requirement of customers are addressed as constraints and are satisfied by the graph decomposition based formulation. The configurations are compared with the baseline (configuration 9) that the personal vehicles are used for the trip. The routing strategy of configurations 2 and 6 is a hybrid routing strategy depends on the occupancy of the vehicles. If the vehicle is occupied, then the shortest time route is used. Otherwise, the eco-route is used.

## V. RESULTS AND DISCUSSION

In the following section, simulation results from the SUMO model are presented. First, we verify that our calibrated simulator can recreate the average speed at evening rush hour of Ann Arbor, and then the model is used to estimate the effect of eco-MOD at city-scale. The fleet size required to serve 4% of the total travel demands for Ann Arbor from 17:00 to 19:00 is estimated using two models. Due to the approximations made by the models, a parametric study of the fleet size is performed using the calibrated traffic simulator to evaluate the system performance. Finally, simulation results of eco-MOD using the configurations from Section IV are presented.

### A. Traffic Simulator Calibration

Assuming that the microscopic driving behavior follows parameters from [45], the demand distribution and route choice are calibrated using data from SPMD. Links with more than 100 events are used for calibration. In the simulation, 150,457 trips are generated from 17:00 to 19:00. The marginal distribution of origins and destinations are shown in Fig. 4 and Fig. 5, with high density indicated by yellow and low density indicated by blue. Measured and simulated average speed normalized using posted speed limit from 17:00 to 17:30 are shown in Fig. 6 and Fig. 7 respectively, with low speed indicated by red and high speed indicated by green, and links without enough data are shown in light gray. The relative error distribution is shown in Fig. 8, with mean relative error equals -1% and the standard deviation equals 25%. As shown in Fig. 6 and Fig. 7, the simulation results show less congestion in the downtown area compared with the measured value, possibly due to our assumption that the flow is only generated by passenger cars, thus the ability to simulate pedestrians and public transits in the downtown is limited. As a result, the extended stops due to pedestrian crossings or bus stops are not captured in the model. However, developing a detailed high fidelity traffic simulator considering multiple categories of traffic participates such as pedestrian and public transit is out of the scope of this study and is left for future works.

When using the simulator to evaluate the Eco-MOD framework, we simulate from 16:00 to 19:00. Only background traffic is generated in the first hour to reach the steady-state of the traffic network. The MOD fleet starts to be deployed in the second hour to reach the steady-state of service fleet. The data from the third hour is used to evaluate the efficiency of the system. The average speed of running vehicles of background traffic simulation is shown in Fig. 9 with the histogram of the average speed at the steady-state shown in yellow. As shown in the figure, the system reaches steady-state within the first hour, and the standard deviation of average speed is 0.18 m/s at the steady-state.

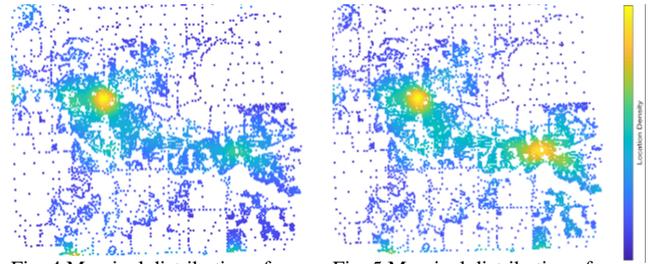

Fig. 4 Marginal distribution of generated trip origins during weekday evening rush hour

Fig. 5 Marginal distribution of generated trip destinations during weekday evening rush hour

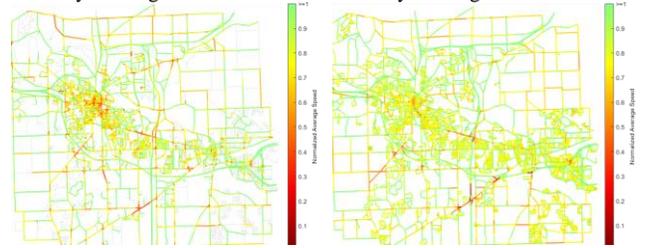

Fig. 6 Measured normalized average speed from 17:00 to 17:30

Fig. 7 Simulated normalized average speed from 17:00 to 17:30

### B. Fleet Size Estimation

To estimate the size of the fleet required to serve the travel demands, we apply the distance-based approach from [46] and the queuing network approach from [47]. We assume that each vehicle only serves one customer for fleet size estimation. Thus the estimation is conservative. However, this doesn't ensure that all travel demands can be served within their time constraints using the algorithm described in Section II. In the algorithm, the idling vehicles are sent to serve the customers whose time constraints cannot be satisfied by the assignment trips, while [47] assumes customers cannot be served will leave

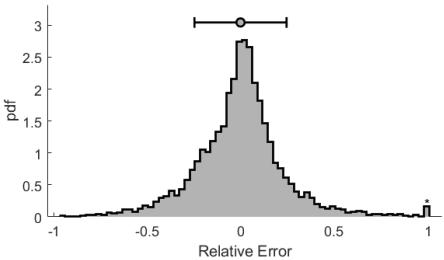

Fig. 8 SUMO simulation average speed relative error distribution

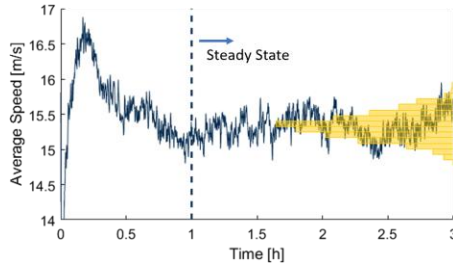

Fig. 9 SUMO simulated network average speed

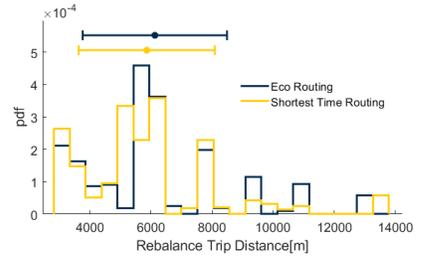

Fig. 10 Rebalance Trip Travel Distance Distribution

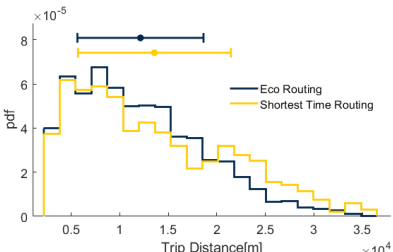

Fig. 11 Generated Trip Travel Distance Distribution

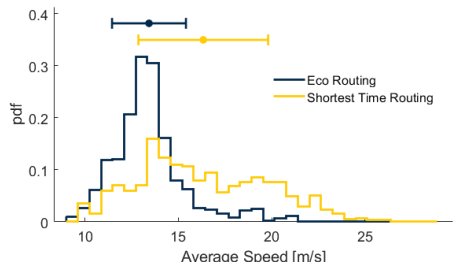

Fig. 12 Average Speed Distribution of Partition Pairs

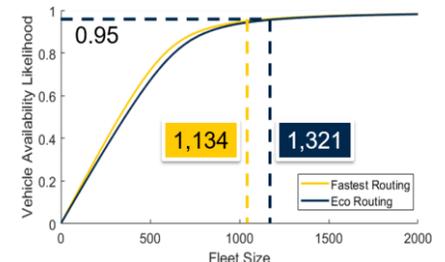

Fig. 13 Vehicle Availability Estimated Using Queuing Network Model

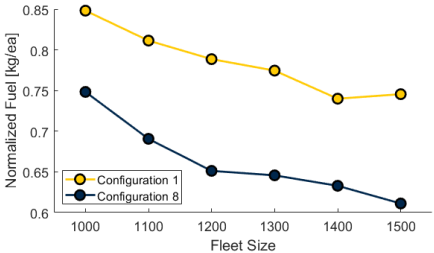
Fig. 14 Fuel consumption normalized with served customer amount

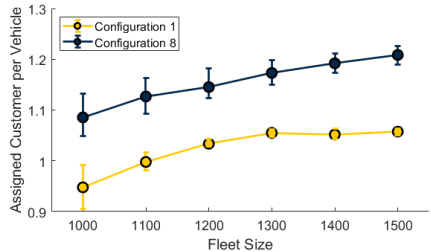
Fig. 15 Average number of assigned customers per running vehicle

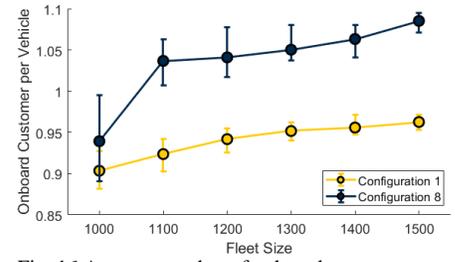
Fig. 16 Average number of onboard customers per running vehicle

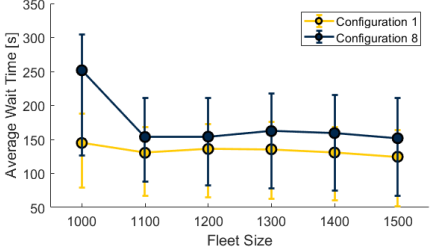
Fig. 17 Average wait time

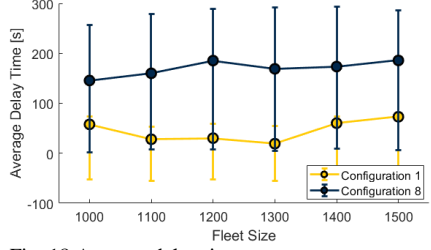
Fig. 18 Average delay time

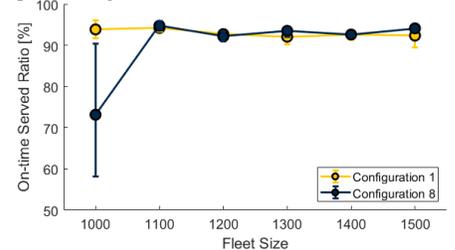
Fig. 19 Ratio of customers served within travel time constraints

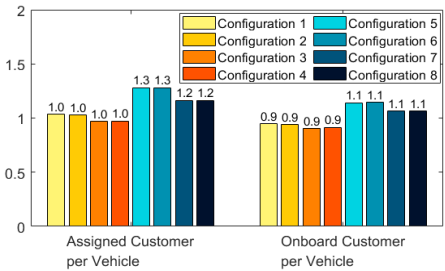
Fig. 20 MOD algorithm performance comparison — average customer assigned and onboard of each vehicle

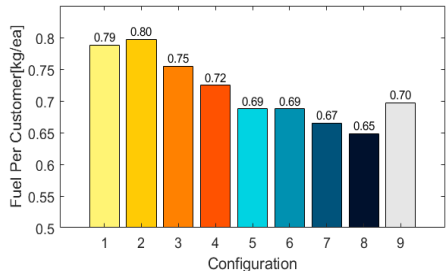
Fig. 21 Fuel Consumption per Customer

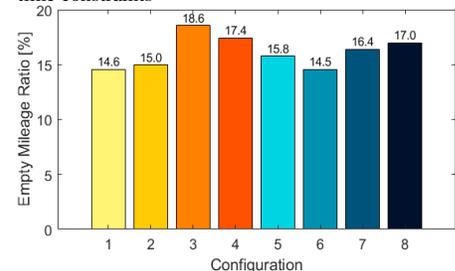
Fig. 22 Empty travel distance ratio

the system instead of waiting for the available vehicle and [46] doesn't take the travel time into consideration. Thus a parametric study is performed to analyze the influence of the fleet size on system performance.

When applying the methods to estimate fleet size, the average travel speed and distance are estimated using the shortest time routing and eco-routing. The distributions of the network statistics required to estimate the fleet size using the distance-based approach are shown from Fig. 10 to Fig. 12. The estimated minimum fleet size for eco-routing is 1,176, and 1,039 for the shortest time routing to serve 4% of the total travel demand from 17:00 to 19:00. Since the approach only addresses the minimum fleet size problem using travel distance and average speed, the wait time of customers can be long [46]. Therefore, the distance-based approach can be used as a lower bound estimation if there is no shared ride.

Availability as a function of fleet size using both shortest time routing and eco-routing is shown in Fig. 13. Due to the lower average speed results from the eco-routing strategy, more vehicles are required to achieve the same availability compared with the shortest time routing. Under the assumptions of queuing network based formulation, to achieve more than 95% availability for all partitions, 1,321 vehicles are required using the eco-routing strategy, and 1,134 vehicles are required using the shortest time routing strategy.

Numerical simulations are used for performance evaluation using different fleet sizes given a max wait time of 5 minutes and a max delay time of 5 minutes for time oriented assignment and fuel oriented assignment (configuration 1 and configuration 8 from TABLE I). The fleet performance is summarized from Fig. 14 to Fig. 19, where 25th and 75th percentiles are represented using error bars. The fuel oriented configuration (configuration 8) consumes less fuel compared with the travel time oriented configuration (configuration 1) as shown in Fig. 14. Due to the fleet cost oriented objective function in the assignment step, the average number of customers per vehicle is higher for configuration 8 (Fig. 15, Fig. 16), indicating more trips are shared. However, the average wait time (Fig. 17) and delay time (Fig. 18) of configuration 8 are longer than configuration 1 for all fleet sizes. For configuration 1, 1,000 vehicles can serve more than 90% of the customers within the time constraints, while 1,100 vehicles are required for configuration 8 (Fig. 19). However, according to Fig. 17, the average wait time reaches steady state with more than 1,100 vehicles for both configurations. In the following section, the fleet size is set to be 1,200.

*C. MOD and Routing Strategy's Influence on Energy*

The main goal of the simulations is to assess the impact of different assignment and routing strategies on the fleet fuel consumption. In this Section, we fix the demand ratio served by the MOD fleet at 4% of the total demand during the weekdays from 17:00 to 19:00. The simulated data from 18:00 ~ 19:00 is used for evaluation after the system reaches steady-state. The fleet size is 1,200, which is necessary to serve 90% of the customers within their travel time constraints for all configurations. The performance of shareability is shown in

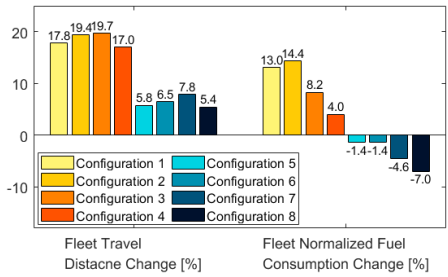

Fig. 23 Change in fleet total travel distance and fuel consumption per customer

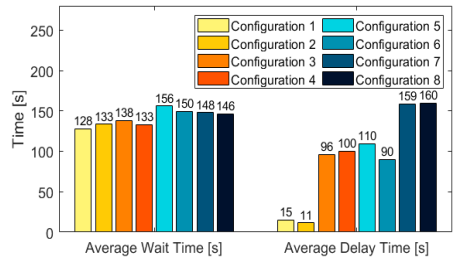

Fig. 24 Time Performance Comparison of Configurations

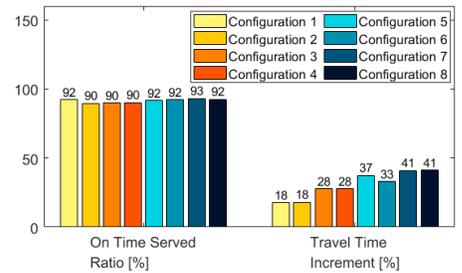

Fig. 25 Ratio of customers served within travel time constraints

Fig. 20. it can be seen that when the fleet cost is minimized, more shared trips are selected, and the average number of assigned customer per vehicle increases from 1.0 to 1.3, and the average number of onboard customers per vehicle increases from 0.9 to 1.1, indicating that more trips are shared and empty vehicle miles is reduced. However, due to the the lower trip average speed, more rebalance trips are assigned for which no shared trips are allowed when eco-routing is applied. The increased amount of the rebalance trips reduced the average number of customers assigned per vehicle from 1.3 to 1.2 and the number of onboard customers from 1.14 to 1.06 when the assignment objective is the fleet fuel consumption.

The performance in fuel consumption and vehicle mileage are summarized in Fig. 21 to Fig. 23. When the objective function of the trip assignment is travel time and the shortest time routing strategy is used, the fuel consumption per customer is increased by 13.0% compared with the baseline when every trip uses a personal vehicle due to the extra empty vehicle mileage for passive and active rebalancing trips while the average number of assigned passengers per vehicle is 1.0. Using eco-routing for trips can reduce fuel consumption, but the normalized fuel consumption is still higher than the baseline. However, if the objective function is to minimize the fleet fuel consumption, with 1.2 to 1.3 passengers assigned to each vehicle on average, the fuel consumption per customer can be reduced by 1.4% to 7.0% compared to the baseline, indicating trip-sharing is the major factor for fuel consumption reduction.

The results indicate that the shared-rides have the potential to reduce the trip fuel consumption by 7%, but if the fleet is not properly operated, the total fuel consumption can increase. The results also indicate that with the same objective function, using eco-routing for trips can further reduce fuel consumption by 8% if the trip assignment objective is travel time, and 5% if the trip assignment objective is fleet fuel consumption compared with the configurations that using the fastest route.

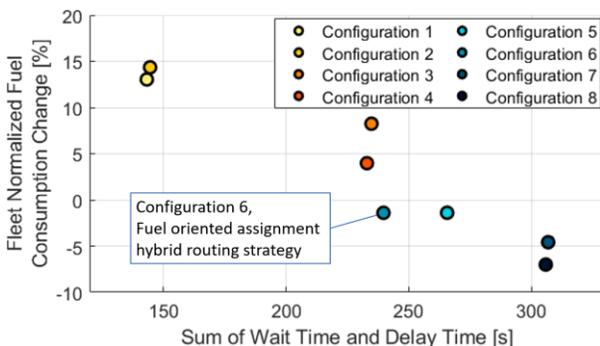

Fig. 26 Tradeoff between time cost defined as sum of average wait time and average delay time, and normalized fuel consumption

The travel time performance is summarized in Fig. 24 and Fig. 25. Since the wait time and delay time of customers are modeled as constraints for trip assignment, all configurations can serve more than 90% of the customers within the travel time constraints. As shown in the plots, shared mobility has the potential to reduce the total fuel consumption but can increase travel time. The objective function can also be defined as a weighted sum of individual benefit and system benefit, and a parametric study can be used to find the Pareto optimal points. The total time cost and fuel consumption for all configurations are summarized in Fig. 26. Configuration 6 (minimizing fleet fuel in assignment and using routing strategy based on vehicle occupancy) shows the best time performance among all fuel oriented configurations and still has the improvement in fleet fuel compared with the baseline as shown in Fig. 26.

VI. CONCLUSIONS

An Eco-MOD fleet assignment framework is developed to minimize fleet fuel consumption while satisfying travel time constraints using a data-driven Bayesian nonparametric fuel consumption model. The system is evaluated using SUMO calibrated with real-world driving data. The algorithm shows the potential to reduce fleet fuel consumption by 7% compared with personal vehicle baseline, while serving more than 90% of the customers within their travel time constraints. The main contributions of this work include: 1) a fuel oriented MOD fleet optimization strategy using data driven fuel consumption model; 2) a traffic simulation framework to verify MOD fleet performance using realistic travel demand; 3) demonstrating the importance of including fuel consumption in the optimization cost function to reduce fleet operating cost.

The analysis assumes that the penetration ratio of MOD fleet is small, thus the influence on the average link travel time is not significant. One potential future direction could be developing a scalable fleet assignment algorithm for large fleet. Also, the fleet is assumed to be homogeneous. Developing fuel model and trip assignment strategy for fleet consisting of multiple vehicle types can also be fruitful.

VII. ACKNOWLEDGMENT

The authors would like to thank the University of Michigan Transportation Research Institute (UMTRI) for making the data from the Safety Pilot Model Deployment (SPMD) project available. This work is sponsored by the U.S. Department of Energy Vehicle Technologies Office under the award DE-EE0007212, an initiative of the Energy Efficient Mobility Systems Program. David Anderson, a Department of Energy

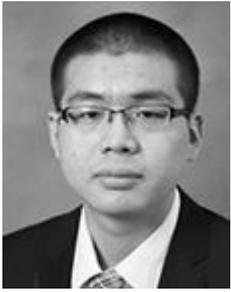

**Xianan Huang** received the Ph.D degree in mechanical engineering from University of Michigan, Ann Arbor, MI, USA, in 2019. From 2013 to 2014 he was an undergraduate researcher at Purdue University. Since 2014 he has been a graduate researcher at University of Michigan, Ann Arbor. His research interests include connected automated vehicle, intelligent transportation system, statistical learning and controls. Dr. Huang's awards and honors include A-Class scholar of Shanghai Jiaotong University and Summer Undergraduate Research Fellowship (Purdue University).

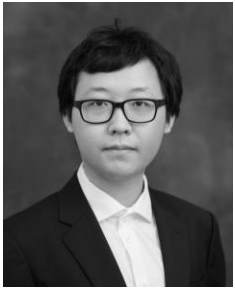

**Boqi Li** received his B.S. degree in Mechanical Engineering at the University of Illinois at Urbana-Champaign. He received his M.S. degree at Stanford University. He is currently a Ph.D. student at the University of Michigan at Ann Arbor. His research interests include eco-routing, reinforcement learning, and cooperative control of connected and automated vehicles.

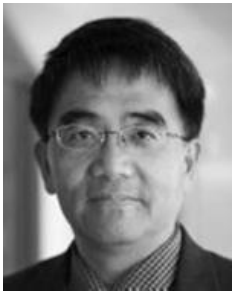

**Huei Peng** received the Ph.D. degree from the University of California, Berkeley, CA, USA, in 1992. He is currently a Professor with the Department of Mechanical Engineering, University of Michigan, Ann Arbor. He is currently the U.S. Director of the Clean Energy Research Center—Clean Vehicle Consortium, which supports 29 research projects related to the development and analysis of clean vehicles in the U.S. and in China. He also leads an education project funded by the Department of Energy to develop ten undergraduate and graduate courses, including three laboratory courses focusing on transportation electrification. He serves as the Director of the University of Michigan Mobility Transformation Center, a center that studies connected and autonomous vehicle technologies and promotes their deployment. He has more than 200 technical publications, including 85 in refereed journals and transactions. His research interests include adaptive control and optimal control, with emphasis on their applications to vehicular and transportation systems. His current research focuses include design and control of electrified vehicles and connected/automated vehicles.

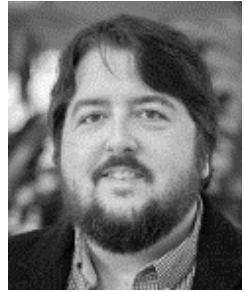

**Joshua Auld** is a Principal Computational Transportation Engineer in Argonne's Vehicle and Mobility Systems Group, and technical manager of transportation systems simulation. He completed his Doctorate in August 2011, in the Civil and Materials Engineering Department of the University of Illinois at Chicago with a concentration in transportation. Dr. Auld has experience in a variety of areas in transportation, with a primary focus on dynamic activity-based travel demand models and the interactions between travel demand and intelligent transportation systems operations. He has over 40 publications in refereed journals and books. He is a member of the TRB Travel Forecasting Resource and Transportation Demand Forecasting Committees.

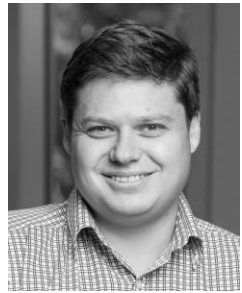

**Vadim Sokolov** received his Ph.D. degree in mathematics from Northern Illinois University in 2008. He is currently an assistant professor in the Systems Engineering and Operations Research Department at George Mason University. He has more than 30 publications in refereed journals and transactions. His research interests include deep learning, Bayesian analysis of time series data, design of computational experiments. Inspired by an interest in urban systems he co-developed mobility simulator called Polaris that is currently used for large scale transportation networks analysis by both local and federal governments. Prior to joining GMU he was a principal computational scientist at Argonne National Laboratory, a fellow at the Computation Institute at the University of Chicago and lecturer at the Master of Science in Analytics program at the University of Chicago.